\documentclass[review]{elsarticle}

\usepackage{lineno}
\modulolinenumbers[5]

\journal{Journal of Sound and Vibration}

\usepackage{amsmath}
\usepackage{epstopdf} 
\usepackage[retainorgcmds]{IEEEtrantools}
\usepackage{mathtools}
\usepackage{siunitx}
\usepackage[colorlinks=true,linkcolor=black,citecolor=black]{hyperref}

\newcommand{\ud}{\mathrm{d}}
\newcommand{\e}{\mathrm{e}}

\newcommand{\sgn}{\mathrm{sgn}}

\renewcommand{\d}{\mathrm{d}}
\renewcommand{\i}{\mathrm{i}}
\newcommand{\TI}{\textrm{TI}}
\newcommand{\ri}{\mathrm{i}}
\graphicspath{{Figs/}{IllustrationFigs/}{DataFigs/}}
\begin{document}

\begin{frontmatter}

\title{Rapid noise prediction models for serrated leading and
trailing edges}

\author[damtp]{Benshuai Lyu\corref{mycorrespondingauthor}}
\cortext[mycorrespondingauthor]{Corresponding author}
\ead{bl362@cam.ac.uk}

\author[damtp]{Lorna J. Ayton}

\address[damtp]{Department of Applied Mathematics and Theoretical Physics,
University of Cambridge, Cambridge CB3 0WA, UK}

\begin{abstract}
    Leading- and trailing-edge serrations have been widely used to reduce the
    leading- and trailing-edge noise in applications such as contra-rotating
    fans and large wind turbines. Recent studies show that these two noise
    problems can be modelled analytically using the Wiener-Hopf method.
    However, the resulting models involve infinite-interval integrals that
    cannot be evaluated analytically, and consequently implementing them poses
    practical difficulty. This paper develops easily-implementable noise
    prediction models for flat plates with serrated leading and trailing edges,
    respectively. By exploiting the fact that high-order modes are cut-off and
    adjacent modes do not interfere in the far field except at sufficiently
    high frequencies, an infinite-interval integral involving two infinite sums
    is approximated by a single straightforward sum. Numerical comparison shows
    that the resulting models serve as excellent approximations to the original
    models. Good agreement is also achieved when the leading-edge model
    predictions are compared with experimental results for sawtooth serrations
    of various root-to-tip amplitudes. Importantly, the models developed in
    this paper can be evaluated robustly in a very efficient manner. For
    example, a typical far-field noise spectrum can be calculated within
    milliseconds for both the trailing- and leading-edge noise models on a
    standard desktop computer. Due to their efficiency and ease of numerical
    implementation, these models are expected to be of particular importance in
    applications where a numerical optimization is likely to be needed.
\end{abstract}


\end{frontmatter}

%\linenumbers

\section{Introduction}
\label{sec:Introduction}
% airfoil noise and then quickly dives into the LE and TE problems
Airfoil noise is important in many applications such as contra-rotating fans
and large wind turbines. It often involves more than one noise generation
mechanism~\cite{Amiet1975,Brooks1989}. Of particular relevance are the
leading-edge (LE) noise and the turbulent boundary-layer trailing-edge (TE)
noise. LE noise is due to the scattering of velocity fluctuations of the
incoming flow by the leading-edge of an airfoil, therefore it is common in
applications with multi-row rotors/stators where the wake flow due the front
row impinges on the downstream blades leading to strong flow-structure
interactions, such as in jet engines and contra-rotating fans. TE noise, on the
other hand, is generated when a turbulent boundary layer convects past and then
gets scattered by the trailing edge of an airfoil~\citep{Howe1978}. It is thus
common in applications with highly turbulent boundary layers, such as wind
turbines.

% leading-edge noise issue 
One of the early research works on LE noise was conducted by \citet{Graham1970},
where similarity rules were established for the unsteady aerodynamic loading of
the airfoil due to sinusoidal gusts at subsonic speed. Following Graham,
\citet{Amiet1975} investigated the acoustic response of an airfoil subject to
sinusoidal incoming gusts. Amiet used the Schwarzschild method and related the
far-field sound Power Spectral Density (PSD) to the wavenumber spectral density
of the vertical velocity fluctuations of the incoming turbulent flow. With an
accurate model for the turbulence wavenumber spectral density, Amiet's approach
has been shown to work well and become an important method for following
studies.
%\citep{Devenport2010}

A serrated LE has been proposed as one of the most promising approaches to
reduce LE
noise~\citep{Bushnell1991,Fish2008,Pedro2008,Hansen2012,Narayanan2015}, and
extensive research has been carried out to study its noise reduction
performance and mechanisms. This includes experimental studies such as those by
\citet{Hansen2012} and \citet{Narayanan2015}, numerical investigations carried
out by \citet{Lau2013}, \citet{Kim2016} and \citet{Turner2016}, and analytical
examinations such as those by \citet{Lyu2017c} and \citet{Ayton2019}.

% trailing-edge noise
Similarly, TE serrations have been widely used as an effective way of reducing
TE noise. A large bulk of literature on this is experimental work. These
include studies by \citet{Dassen1996,
Chong2013a,Moreau2013,Oerlemans2009,Gruber2013,Chong2015,Leon2017}, etc.
Numerical techniques have also been widely used to study the TE serrations as a
way of reducing TE noise, see for example those by
\citet{Jones2012,Sanjose2014} and \citet{Van2017}. A number of authors have
also conducted analytical studies. Some of the early analytical works include
those by \citet{Howe1991a,Howe1991b}, where a tailored Green's function was
used to predict the far-field sound generated by flat plates with sinusoidal
and sawtooth serrations, respectively. However, Howe's model dramatically
overpredicted the sound reduction achieved by using TE serrations. Howe's
approach was later used by \citet{Azarpeyvand2013} to study the noise reduction
characteristics of other serration geometries. The recent work by
\citet{Lyu2016a,Lyu2015}, on the other hand, used Amiet's approach and extended
Amiet's model~\citep{Amiet1976b,Sinayoko2014} for a straight trailing edge to
the serrated case. The results showed that the principal noise reduction
mechanism was due to the destructive interference and the predicted noise
reduction was more realistic compared to experimental results.

%Introcue the very reccentw work similarity between the two hence serrations
%are used in both cases. establish research gaps
Although the TE noise and LE noise are due to different noise generation
mechanisms, mathematically they bear a striking similarity, hence, the
techniques used to model the two problems are expected be similar to each
other. For example, recent work \citep{Ayton2018d,Ayton2019} shows that both
the serrated LE noise and TE noise can be modelled analytically using the
Wiener-Hopf method. This approach has shown good agreement with experiments for
LE noise~\cite{Ayton2019}. However, both the LE and TE solutions involve an
infinite-interval integral, which makes their implementations both difficult
and error-prone.

% paper structure.
In this paper, we address this issue. By exploiting the fact that high-order
modes are cut-off and little coupling between expanded modes occurs except at
very high frequencies, we replace the infinite-interval integral that involves
two infinite sums with one straightforward sum. The simplified model takes a
particularly concise form when the serration wavelength is small compared to
the transformed acoustic wavelength. The final results can therefore be easily
implemented numerically in a robust and efficient manner.

This paper is structured as follows. Section~\ref{sec:Analysis} shows the
essential analytical steps to reach the final results for both the LE and TE
noise problems, respectively. Section~\ref{sec:Comparision} presents a
comparison between the approximated results obtained in this paper and those
obtained from the full analytical solutions. The following section uses the
simplified leading-edge model to compare with the leading-edge noise spectra
observed in experiments. The final section concludes this paper and lists
directions for future work.

% leading-edge noise research evolmen% leanding edge noise research gap
% similarity of LE and TE problems.
% what we do in this paper and paper structure

\section{The leading-edge and trailing-edge noise models}
\label{sec:Analysis}
\begin{figure}
    \centering
    \includegraphics[width=0.95\textwidth]{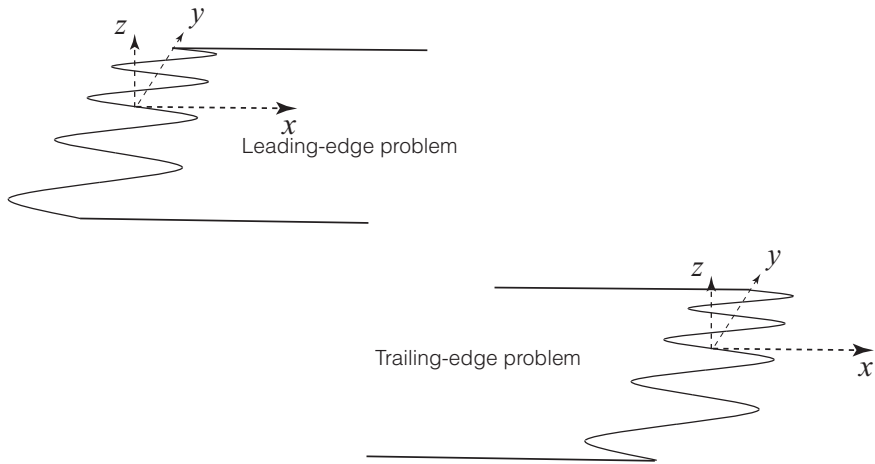}
    \caption{Schematic illustrations of the simplified leading-edge and
    trailing-edge noise problems. The two problems bear a striking similarity
between each other.}
    \label{fig:schematic}
\end{figure}

% similarity and introduce general setup introducing mean flow rho U and
% non-dimensionalizing strategy
As mentioned in Section~\ref{sec:Introduction}, the TE and LE noise problems
bear a striking similarity between each other. In either case, to allow the
analytical derivation to continue, the serrated airfoil is often assumed to be
a semi-infinite plate~\citep{Amiet1975,Roger2005,Ayton2019,Lyu2017c,Lyu2016a}
placed in a uniform incoming flow of constant density $\tilde{\rho}$ and
velocity $\tilde{U}$ at zero angle of attack, as shown in
figure~\ref{fig:schematic}. The speed of sound is denoted by $\tilde{c}_0$. In
the rest of this paper, the serration wavelength is used to normalized the
length dimension, while $\tilde{\rho}$ and $\tilde{U}$ are used to
non-dimensionalize other dynamic variables such as the velocity potential and
pressure. In this paper, we restrict our attention to periodic leading-edge and
trailing-edge serrations. Because the geometric parameters are normalized by
the serration wavelength, the serrations have a period $1$. The normalized
root-to-tip length is given by $2h$. Let $x$, $y$, $z$ denote the streamwise,
spanwise and normal to the plate directions, respectively, and  the coordinate
origin is fixed in the middle between the root and tip. In such a coordinate
frame, the serration profile can be described by $x = hF(y)$, where $F(y)$ is a
single-valued function that has a maximum value $1$ and minimum value $-1$.
Moreover, we require $1$ to be the smallest period. Other than these
constraints, the function $F(y)$ is arbitrary.

% introduce difference
Figure~\ref{fig:schematic} shows the similarity between the simplified
leading-edge and trailing-edge noise problems due to coordinate symmetry.
Despite this symmetry, the physics they represent is quite different. For the
leading-edge problem, the unsteady flow fluctuations, due to the incoming
turbulence convected by the mean flow, are scattered into sound near the
leading edge of the flat plate, whereas in the trailing-edge problem the source
of scattering is the turbulence beneath turbulent boundary layers. The boundary
conditions required by these two problems are therefore quite different. As
such, we need to discuss them separately. 

% introduce a single serration

\subsection{The leading-edge noise problem}
\label{subsec:leadingedgenoiseproblem}
% introduce the physics to math problem
When the turbulence in the mean flow passes the leading edge, a scattered
potential flow is induced. The scattered potential ensures that appropriate
boundary conditions are satisfied. In the leading-edge noise problem, the
vertical velocity fluctuation of the incoming turbulence is of primary concern.
The turbulence in the mean flow consists of a wide range of time and length
scales. However, one can always perform a Fourier Transformation on the
incoming vertical velocity field, such that it can be written as
\begin{equation}
    w_i = \int_{-\infty}^\infty \hat{w}_0(\omega, k_2)
    \e^{\i(-\omega t + k_1 x + k_2 y)} \ud k_2,
    \label{equ:gust}
\end{equation}
where $t$ denotes time, $\hat{w}_0$ the velocity fluctuation in the $z$
direction, $\omega$ the angular frequency and $k_1$ and $k_2$ the wavenumbers
in the streamwise and spanwise directions, respectively. The turbulence is
assumed to be frozen and convects downstream at a non-dimensional speed of $1$.
Therefore, one has $k_1 = \omega$.

Let $\phi_s$ denote this scattered velocity potential. One can show that
$\phi_s$ satisfies the convective wave equation
\begin{equation}
    \nabla^2 \phi_s - M^2 \left(\frac{\partial}{\partial t} %
    + \frac{\partial}{\partial x} \right)^2 \phi_s = 0,
    \label{equ:convectiveWaveEquation}
\end{equation}
where $M = \tilde{U}/\tilde{c}_0$. To ensure that the normal velocity on the
plate vanishes, we require
\begin{equation}
    \left.\frac{\partial \phi_s}{\partial z}\right|_{z=0}%
    = - w_i, \quad x > hF(y).
    \label{equ:bc1}
\end{equation}
The scattering problem is anti-symmetric across $z = 0$, therefore we also have
\begin{equation}
    \left.\phi_s\right|_{z=0} = 0, \quad x < hF(y).
    \label{equ:bc2}
\end{equation}

This mixed boundary condition problem can be solved using the Wiener-Hopf
method~\citep{Ayton2019,Lyu2018c}. For the sake of brevity we omit the details
of the solving procedure. Interested readers are referred to \citet{Ayton2019}
and the appendix of \citet{Lyu2018c}. Here we only give the results in the
acoustical far-field as
\begin{equation}
    p(\omega, r, \theta, y) \approx \int_{-\infty}^\infty 
    H_l(\omega, \boldsymbol{x}, k_2) \hat{w}_0(\omega, k_2) \d k_2,
    \label{equ:farfieldPressureTotal} 
\end{equation}
where 
\begin{equation}
    \begin{aligned}
	H_l(\omega, \boldsymbol{x}, k_2) = \frac{\e^{\i \pi/4}}{\sqrt{\pi}} 
	& \e^{-\i k M x /\beta^2} \cos\frac{\theta}{2}\\
	& \sum_{n=-\infty}^{\infty} 
	\frac{\frac{k_1}{\beta^2} - \kappa_n \cos\theta}
	{\bar{k}_1 - \kappa_n \cos\theta}
	\frac{1}{\sqrt{\bar{k}_1 + \kappa_n}}
	\frac{\e^{\i \kappa_n r}}{\sqrt{r}} \e^{\i \chi_n y}
	E_n(-\kappa_n \cos\theta).
    \end{aligned}
    \label{equ:transferFunction}
\end{equation}
In equation~\ref{equ:transferFunction}, $r$, $\theta$ and $y$ denote,
respectively, the radial, polar and axial axes of the stretched cylindrical
coordinate system $(x/\beta, y, z)$, i.e. $y$ denotes the axial axis and
$\theta$ is the polar angle to the stretched axis $x/\beta$ in the $(x/\beta,
z)$ plane ($\theta=0$ corresponds to the $x/\beta$ axis) and $r =
\sqrt{(x/\beta)^2 + z^2}$. In addition, one has $k = \omega M$, $\beta =
\sqrt{1 - M^2}$, $\bar{k}_1 = k_1 / \beta$, $\chi_n = k_2 + 2 n \pi$, $\kappa_n
= \sqrt{k^2 - \chi_n^2}$ and
\begin{equation}
    E_n(-\kappa_n \cos \theta) = \int_0^1 \e^{\i (\bar{k}_1 - \kappa_n
    \cos\theta) \bar{h} F(\eta)} \e^{-\i 2 n \pi \eta} \ud \eta,
    \label{equ:EnFunction}
\end{equation}
where $\bar{h} = h /\beta$.

Since the incoming turbulence is statistically stationary, the far-field sound
is best formulated statistically. Routine procedure shows that the far-field
sound PSD is given by
\begin{equation}
    \Psi(\omega, r, \theta, y) = \lim_{T\to\infty} \frac{\pi}{T} 
    p(\omega, r, \theta, y) p^\ast(\omega, r, \theta, y),
    \label{equ:psdDefinition}
\end{equation}
where $2T$ is the time interval used to performed temporal Fourier
Transformation to obtain $p$ and the asterisk denotes the complex conjugate.
Substituting equations~\ref{equ:farfieldPressureTotal} into
\ref{equ:psdDefinition}, we can show that
\begin{equation}
    \begin{aligned}
	\Psi(\omega, r, \theta, y) \approx &
	\frac{1}{\pi r} \cos^2 \frac{\theta}{2}\\
	&\times \int_{-\infty}^\infty 
	\Pi_l(\omega, k_2)
	\sum_{n=-\infty}^{\infty}
	\frac{\frac{k_1}{\beta^2} - \kappa_n \cos\theta}
	{\bar{k}_1 - \kappa_n \cos\theta}
	\frac{E_n(-\kappa_n\cos\theta)}
	{ \sqrt{\bar{k}_1 + \kappa_n}} 
	\e^{\ri \chi_n y}
	\e^{\ri \kappa_n r}
	\\
	&\times 
	\sum_{m=-\infty}^{\infty}
	\left[
	    \frac{\frac{k_1}{\beta^2} - \kappa_m \cos\theta}
	    {\bar{k}_1 - \kappa_m \cos\theta}
	    \frac{E_m(-\kappa_m\cos\theta)}
	    {\sqrt{\bar{k}_1 + \kappa_m}} 
	    \e^{\ri \chi_m y}
	    \e^{\ri \kappa_m r}
	\right]^\ast \ud k_2,
    \end{aligned}
    \label{equ:psdFarfield}
\end{equation}
where $\Pi_l(\omega, k_2)$ is the wavenumber frequency spectrum of the vertical
velocity fluctuations due to the incoming turbulence.

Experiments and theories have shown that narrow serrations (a small serration
wavelength) are more effective than wide serrations in reducing the LE
noise~\citep{Narayanan2015,Lyu2017c}. This is related to the spanwise
correlation length of the incoming gust, hence to the integral scale of the
incoming turbulence. A detailed discussion was given by \citet{Lyu2017c} and
\citet{Lyu2018c}. Consequently, for practical usage, we only need to restrict
our attention to narrow serrations. Under the assumption of narrow
serrations, equation~\ref{equ:psdFarfield} can be further simplified. First, we
only need to investigate the case where both $\kappa_n$ and $\kappa_m$ are
real, because otherwise the exponential term $\e^{\ri \kappa_n r}(\e^{\ri
\kappa_m r})^\ast$ causes the whole term to decay exponentially in the far
field. Noting that $\kappa_n = \sqrt{\bar{k}^2 - \chi_n^2}$, we see that
$\kappa_n$ is real only when $-\bar{k} < \chi_n < \bar{k}$. Because we restrict
to the case where the serration wavelength is small, in the frequency range of
interest we may have $\bar{k} < \pi$. It is, therefore, permissible to have
\begin{equation}
    \e^{\ri \kappa_n r}(\e^{\ri \kappa_m r})^\ast = \delta_{nm}
\sgn(\Re(\kappa_n)),
    \label{equ:deltanm}
\end{equation}
where $\sgn(0) = 0$ and $\sgn(x) = \pm 1$ when $\pm x >0$. Note
equation~\ref{equ:deltanm} shows that its right hand side vanishes when
$\kappa_n$ is imaginary. This implies that
\begin{equation}
    \begin{aligned}
	\Psi(r, \theta, y) \sim & \frac{1}{\pi r}
	\cos^2\frac{\theta}{2}
	\sum_{n=-\infty}^{\infty} \int_{-\infty}^\infty \Pi_l(\omega, k_2)\\
	&\times \left| (\frac{k_1}{\beta^2} - \kappa_n \cos\theta)
	\frac{E_n(-\kappa_n\cos\theta)}
	{(-\kappa_n\cos\theta +\bar{k}_1) \sqrt{\bar{k}_1 + \kappa_n}} 
	\right|^2 
	\sgn(\Re(\kappa_n)) \ud k_2.
    \end{aligned}
    \label{equ:psdFarfieldSimplified1}
\end{equation}

Second, in light of the fact that $\kappa_n$ is real only when $-\bar{k} <
\chi_n < \bar{k}$ and the serration wavelength is small, the integrand does not
vanish only when $-2n\pi-\bar{k} \le k_2 \le -2n\pi + \bar{k}$. Over such a
typically small range of $k_2$, the integrand of
equation~\ref{equ:farfieldPressureTotal} does not vary significantly due to its
algebraic dependence on $k_2$ (provided the Mach number is not close to $1$).
Hence we can factor the $k_2$ dependence out of the integral, and change the
integration interval to $-2n\pi-\bar{k}$ to $-2n\pi + \bar{k}$, without causing
significant errors. Upon doing so, equation~\ref{equ:farfieldPressureTotal}
simplifies to 
\begin{equation}
    \begin{aligned}
	\Psi(r, \theta, y) \sim 
	\frac{2\bar{k}}{\pi r}\cos^2\frac{\theta}{2} 
	\frac{(\frac{k_1}{\beta^2} - \bar{k}
	\cos\theta)^2}
	{(\bar{k}_1 - \bar{k}\cos\theta )^2 (\bar{k}_1 + \bar{k})}
	\sum_{n=-\infty}^{\infty} \Pi_l(\omega, 2n\pi)
	\left|E_n(-\bar{k}\cos\theta) \right|^2.
    \end{aligned}
    \label{equ:psdFarfieldSimplified2}
\end{equation}

Equation~\ref{equ:psdFarfieldSimplified2} is obtained by assuming the
serrations are narrow. It would of course be useful to know how narrow can be
regarded as appropriate. This can be obtained from the criterion that $\bar{k}
< \pi$ (recall lengths are non-dimensionlised by the serration wavelength). In
fact, when $\pi < \bar{k}\le2\pi$, the overlap between adjacent modes is still
rather weak, therefore it is often permissible to assume that the approximation
is still valid when $\bar{k}\le 2\pi$. It is clear that this inequality depends
only on the (dimensional) serration wavelength, (dimensional) acoustic
wavenumber and Mach number. This is likely to be satisfied in practical
applications. To put this into perspective, let us take an typical example
applicable in the wind industry for an airfoil of chord $1$~m placed in a mean
flow of Mach number $0.2$. The serration wavelength is around $2$ cm while the
serration root-to-tip is around $10$ cm. The inequality will therefore hold for
a frequency up to $17$ KHz, which is near the upper limit of the audible
frequency range. Thus, our approximation is valid for the full range of
practical interest in this case.

\subsection{The trailing-edge noise problem}
\label{subsec:trailingedgenoiseproblem}
When the turbulent boundary layer convects past the trailing edge of a flat
plate, a scattered pressure field is induced. In a similar manner, we may write
the wavenumber frequency spectrum of the wall pressure fluctuations beneath the
boundary layer as 
\begin{equation}
    p_i = \int_{-\infty}^\infty \hat{p}_0(\omega, k_2) 
    \e^{\i (-\omega t + k_1 x + k_2 y)} \ud k_2,
    \label{equ:pressureSpectrum}
\end{equation}
where relevant quantities are defined in a similar way as those defined in
section~\ref{subsec:leadingedgenoiseproblem}, except here $\hat{p}_0$ is the
amplitude of the Fourier component of wall pressure fluctuations and $k_1 =
\omega / \alpha$, where $\alpha$ is a constant. In other words, these pressure
fluctuations are assumed to convect at a speed of $\alpha$. Here we use a
typical value of $\alpha \approx 0.7$~\citep{Lyu2016a}. 
% we need to introduce the cylindrical coordinate earlier

Let $p_s$ denote the scattered pressure field, which satisfies the convective
wave equation, i.e.
\begin{equation}
    \nabla^2 p_s - M^2 \left(\frac{\partial}{\partial t} %
    + \frac{\partial}{\partial x} \right)^2 p_s = 0.
    \label{equ:convectiveWaveEquationTE}
\end{equation}
The boundary conditions are such that the normal velocity on the plate
vanishes, i.e.
\begin{equation}
    \left.\frac{\partial p_s}{\partial z}\right|_{z = 0} = 0, \quad x < hF(y),
    \label{equ:TEBC1}
\end{equation}
and that the scattered pressure is $0$ on the semi-infinite plane $z =0$ and $x
> h F(y)$, i.e.
\begin{equation}
   \Delta p_s|_{z = 0} = - p_i, \quad x > h F(y),
    \label{equ:TEBC2}
\end{equation}
where $\Delta p_s$ denotes the pressure jump across the plate.
The solution $p_s$ satisfying equation~\ref{equ:convectiveWaveEquation} subject
to the boundary conditions shown in equations~\ref{equ:TEBC1} and
\ref{equ:TEBC2} can be found (see for example \citet{Ayton2018d}) to be
\begin{equation}
    p_s(\omega, r, \theta, y) \approx \int_{-\infty}^\infty 
    H_t(\omega, \boldsymbol{x}, k_2) \hat{p}_0(\omega, k_2) \d k_2,
    \label{equ:farfieldPressureTotalTE} 
\end{equation}
where 
\begin{equation}
    \begin{aligned}
	H_t(\omega, \boldsymbol{x}, k_2) = \frac{\e^{\i \pi/4}}{\sqrt{\pi}} 
	& \e^{-\i k M x /\beta^2} \sin\frac{\theta}{2}\\
	& \times \sum_{n=-\infty}^{\infty} 
	\frac{\sqrt{-\bar{k}_1 - \kappa_n}}
	{2\ri(\bar{k}_1 - \kappa_n \cos\theta)}
	\frac{\e^{\i \kappa_n r}}{\sqrt{r}} \e^{\i \chi_n y}
	E_n(-\kappa_n \cos\theta),
    \end{aligned}
    \label{equ:transferFunctionTE}
\end{equation}
where $r$, $\theta$ are defined similar to those shown in
Section~\ref{subsec:leadingedgenoiseproblem}. In addition, $\chi_n$ and
$\kappa_n$ are defined the same as those in
Section~\ref{subsec:leadingedgenoiseproblem}. However, we now define $\bar{k}_1
= (k_1 + (k M - k_1 M^2))/\beta$ and 
\begin{equation}
    E_n(-\kappa_n \cos \theta) = \int_0^1 \e^{\i (\bar{k}_1 - \kappa_n
    \cos\theta) \bar{h} F(\eta)} \e^{-\i 2 n \pi \eta} \ud \eta,
\end{equation}
where $\bar{h}$ is similarly defined as $h /\beta$. Note here that the
definitions of $k_1$ and $\bar{k}_1$ in this trailing-edge noise problem are
different from those in the leading-edge noise problem.

In a very similar manner, the
far-field sound PSD can be approximated, upon assuming the serration wavelength
is sufficiently small such that $\bar{k} < \pi$, to be
\begin{equation}
    \begin{aligned}
	\Psi&(r, \theta, y) \sim  \frac{1}{4 \pi r}
	\sin^2\frac{\theta}{2}\\
	&\times \sum_{n=-\infty}^{\infty} \int_{-\infty}^\infty \Pi_t(\omega, k_2)
	\left| 
	\frac{\sqrt{-\bar{k}_1 - \kappa_n}}
	{(\bar{k}_1 - \kappa_n \cos\theta)}
	E_n(-\kappa_n \cos\theta)
	\right|^2  \sgn(\Re(\kappa_n))\ud k_2,
    \label{equ:psdFarfieldSimplifiedTE}
\end{aligned}
\end{equation}
where $\Pi_t(\omega, k_2)$ denotes the wall pressure fluctuations wavenumber
frequency spectrum beneath the turbulent boundary layer close to the trailing
edge. Equation~\ref{equ:psdFarfieldSimplifiedTE} can be further simplified by
replacing the integral with a simple sum to be
\begin{equation}
    \Psi(r, \theta, y) \sim \frac{\bar{k}}{2 \pi r}
	\sin^2\frac{\theta}{2}
	\frac{\bar{k}_1 + \bar{k}}
	{(\bar{k}_1 - \bar{k}\cos\theta)^2}
	\sum_{n=-\infty}^{\infty} \Pi_t(\omega, 2n\pi)
	\left|E_n(-\bar{k}\cos\theta)
	\right|^2 .
    \label{equ:psdFarfieldFinalTE}
\end{equation}
It is worth noting equation~\ref{equ:psdFarfieldFinalTE} bears a striking
similarity to equation~2.59 in the work of \citet{Lyu2016a}. The fact that two
completely different approaches lead to consistent results of the same form
shows that the essential physics are captured in both models. These two
equations show that higher-order modes are still cut-on and contribute to the
far-field (the earlier argument in \citet{Ayton2018d} that higher-order modes
were neglected in the model of \citet{Lyu2016a} was not an accurate statement).

\section{Comparison with exact solutions}
\label{sec:Comparision}
In Section~\ref{sec:Analysis}, we reduce the complex original model to a
straightforward sum and simplify the result significantly when the serrations
are sufficiently narrow (i.e. serration wavelength is small compared to the
transformed acoustic wavelength). In this section, to assess how accurate the
approximations are, we perform a direction comparison between the full and the
simplified solutions. Firstly, we choose to compare the solutions for LE
serrations of a sawtooth profile as an illustration.

\subsection{The leading-edge noise problem}
\label{subsec:LEComparison}
To enable this comparison, we need a realistic wavenumber spectrum
$\Pi_l(\omega, k_2)$ to model the incoming turbulence. There are a number of
empirical models available. As an illustration, we use the one developed from
Von K\'{a}rm\`{a}n spectrum. Based on this, it can be shown that $\Pi_l(\omega,
k_2)$, i.e. the wavenumber frequency spectrum of the incoming vertical
fluctuation velocity, can be written as~\citep{Amiet1975,Lyu2016b,Lyu2017c}
\begin{equation}
    \Pi_l(\omega, k_2) = \frac{4\TI^2}{9\pi k_e^2} 
    \frac{\hat{k}_1^2 + \hat{k}_2^2}{(1 + \hat{k}_1^2 + \hat{k}_2^2)^{7/3}},
    \label{equ:vonKarmanSpectrum}
\end{equation}
where $\TI$ denotes the turbulent intensity and $k_e$, $\hat{k}_1$ and
$\hat{k}_2$ are given by
\begin{equation}
    k_e = \frac{\sqrt{\pi}\Gamma(5/6)}{L_t \Gamma(1/3)},\quad 
    \hat{k}_1 = \frac{k_1}{k_e}, \quad
    \hat{k}_2 = \frac{k_2}{k_e}.
\end{equation}
In the above equations, $L_t$ is the integral scale of the turbulence (also
normalized by the serration wavelength) and $\Gamma(x)$ is the Gamma function.
\begin{figure}
    \centering
    \includegraphics[width=0.8\textwidth]{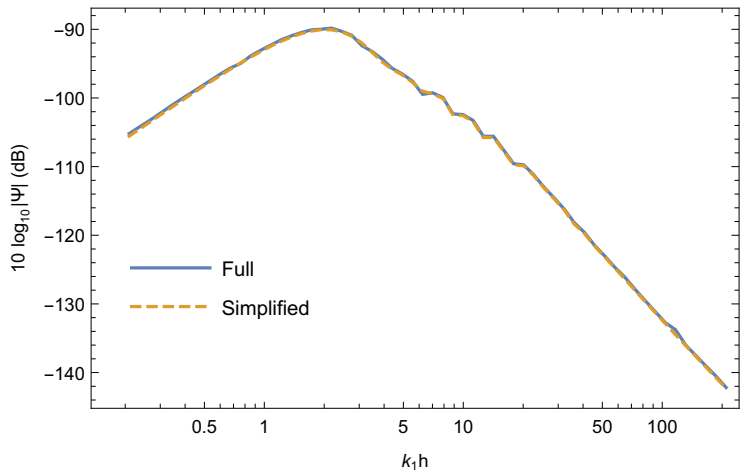}
    \caption{Comparison of predicted LE noise spectra from the full and
    simplified models: $M = 0.18$, $\mathrm{TI} = 0.025$, $h = 5$, $L_t =0.3$,
$r = 30$, $\theta = 90^\circ$ and $y = 0$.}
    \label{fig:validation}
\end{figure}

In order to put equation~\ref{equ:vonKarmanSpectrum} into perspective, we
require a realistic set of physical parameters of the incoming flow. For the
sake of convenience, we use those given in previous
experiments~\citep{Narayanan2015,Lyu2017c}, where $M  = 0.18$, $\TI = 0.025$
and $L_t = 0.3$. As an illustrative example, we use sawtooth serrations with $h
= 5$. The observer distance is fixed at $r= 30$ in the plane of $y = 0$, but
the observer angle is varied from $\theta = 90^\circ$ to $20^\circ$. The
far-field PSDs are evaluated from equations~\ref{equ:psdFarfield} and
\ref{equ:psdFarfieldSimplified2}, respectively. 

The comparison of the predicted noise spectra at $\theta = 90^\circ$ is shown
in figure~\ref{fig:validation}. The solid line is obtained from the full
solution, i.e. equation~\ref{equ:psdFarfield}, while the dashed line is from
the simplified solution, i.e. equation~\ref{equ:psdFarfieldSimplified2}. It is
clear that the two solutions agree excellently well with each other over the
entire frequency range of interest. We choose $h = 5$ because this represents
sharp serrations where the approximation is the least accurate. At such a large
value of $h$ we can hardly see the difference between the full and simplified
solutions. We can therefore expect at least similar, if not better, agreement
for smaller values of $h$. 

\begin{figure}
    \centering
    \includegraphics[width=0.8\textwidth]{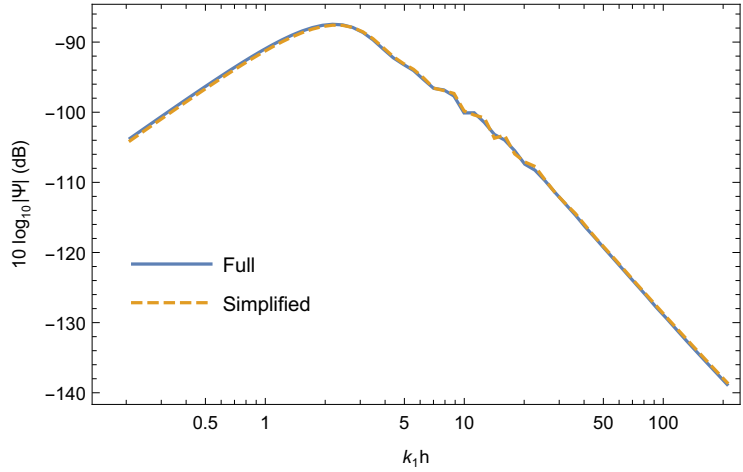}
    \caption{Comparison of predicted LE noise spectra from the full and
    simplified models: $M = 0.18$, $\mathrm{TI} = 0.025$, $h = 5$, $L_t =0.3$,
$r = 30$, $\theta = 45^\circ$ and $y = 0$.}
    \label{fig:LEComparison45}
\end{figure}

\begin{figure}
    \centering
    \includegraphics[width=0.8\textwidth]{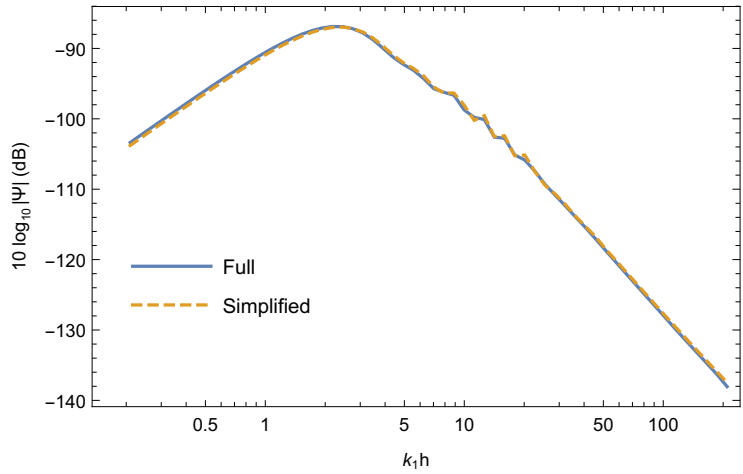}
    \caption{Comparison of predicted LE noise spectra from the full and
    simplified models: $M = 0.18$, $\mathrm{TI} = 0.025$, $h = 5$, $L_t =0.3$,
$r = 30$, $\theta = 20^\circ$ and $y = 0$.}
    \label{fig:LEComparison20}
\end{figure}

Figure~\ref{fig:validation} is for a fixed observer at $90^\circ$ above the
leading edge. Figure~\ref{fig:LEComparison45} shows the predicted noise
spectra for the observer at $\theta = 45^\circ$. The agreement is similar to
that shown in figure~\ref{fig:validation}, and the simplified model serves as
an excellent approximation to the fully integrated solution.
Figure~\ref{fig:LEComparison20} shows the simplified and full spectra when
$\theta = 20^\circ$, and the agreement continues to be very good.

It is worth mentioning, however, the computational costs are very different for
the two solutions. For the full integral solution given by
equation~\ref{equ:psdFarfield}, it takes around one hour and a half to obtain
the noise spectrum at a single observer location, whereas on average only $5$
ms is needed for the simplified model given by
equation~\ref{equ:psdFarfieldSimplified2}. The simplified model is faster than
the original model by a factor of around $720,000$. More importantly, since no
numerical integration of irregular integrands is involved, the computation is
much more robust.

\subsection{The trailing-edge noise problem}
\label{subsec:TEComarison}
We now compare the approximated model to the full model for the TE noise
problem. Similarly, the wall pressure wavenumber spectrum needs to be modelled.
As an illustrative example, it suffices to choose Chase's
model~\cite{Chase1987,Lyu2016a}, i.e. 
\begin{equation}
    \Pi_t(\omega, k_2) = \frac{4 C_m v^{\ast 4} \delta^4 k_1^2}{\alpha
    \left[(k_1^2 + k_2^2)\delta^2 + \chi^2\right]^2},
\end{equation}
where $C_m \approx 0.1533$, $v^\ast \approx 0.03$, $\chi \approx 1.33$, and
$\delta$ denotes the non-dimensional boundary layer thickness. In this paper,
we let $\delta$ take an approximate value of $1.01$, which corresponds to a
realistic non-dimensional boundary layer thickness for a dimensional chord
length of $1$ m when the serration wavelength is $0.02$ m.

To enable a direct comparison between the approximated and full results, we
again use a sawtooth serration profile with $h = 5$. The observer distance is
fixed to be $r = 30$ in the plane of $y = 0$, and the observer angle is varied
from $\theta = 90^\circ$ to $20^\circ$. The predicted far-field spectra are
plotted using the full solution based on
equation~\ref{equ:farfieldPressureTotalTE} and the approximated solution shown
in equation~\ref{equ:psdFarfieldFinalTE}, respectively.
 
\begin{figure}
    \centering
    \includegraphics[width=0.8\linewidth]{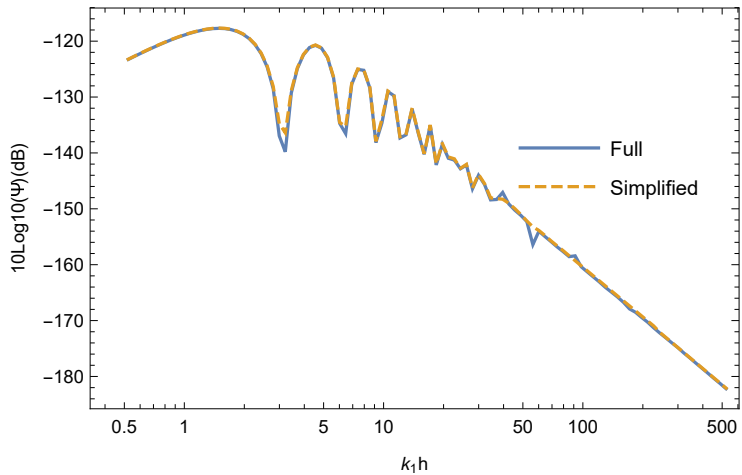}
    \caption{Comparison of predicted TE noise spectra from the full and
    simplified models: $M = 0.1$, $h = 5$, $\delta = 1.01$, $r = 30$, $\theta =
90^\circ$ and $y = 0$.}
    \label{fig:TEComparison90}
\end{figure}
The comparison of the noise spectra at $\theta = 90^\circ$ is shown in
figure~\ref{fig:TEComparison90}. As we can see the approximated solution agrees
with the full solution with virtually no difference over the entire frequency
range. Note that when $k_1 h$ is close to $500$, $\bar{k}$ is slightly larger
than $2\pi$. But the difference between the two lines is still hardly
observable. Therefore, the condition $\bar{k} < 2\pi$, although likely to be
satisfied in most practical cases, may be further relaxed in practice. 

\begin{figure}
    \centering
    \includegraphics[width=0.8\linewidth]{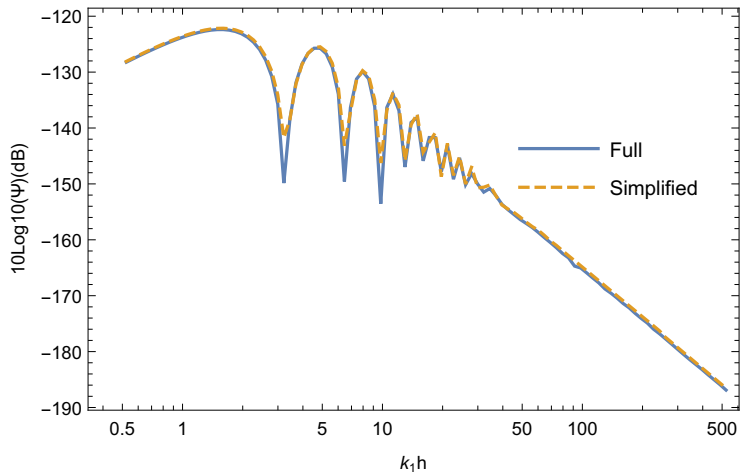}
    \caption{Comparison of predicted TE noise spectra from the full and
    simplified models: $M = 0.1$, $h = 5$, $\delta = 1.01$, $r = 30$, $\theta =
45^\circ$ and $y = 0$.}
    \label{fig:TEComparison45}
\end{figure}
Figure~\ref{fig:TEComparison90} shows the comparison of the predicted spectra
for $\theta = 45^\circ$. The agreement continues to be very good, except slight
disagreement occurring near the minima of the oscillation frequencies. The
strong oscillations predicted by the models are due to the large value of $h$,
i.e. the use of sharp serrations, leading to strong destructive interference
(in experiments, however, these large dips are unlikely to be observed since
the turbulence within the boundary layer is not strictly frozen). We choose
this large value to examine how the simplified model works in the least
accurate case. Had we used smaller values of $h$, these oscillations would have
disappeared~\cite{Ayton2018d}.
\begin{figure}
    \centering
    \includegraphics[width=0.8\linewidth]{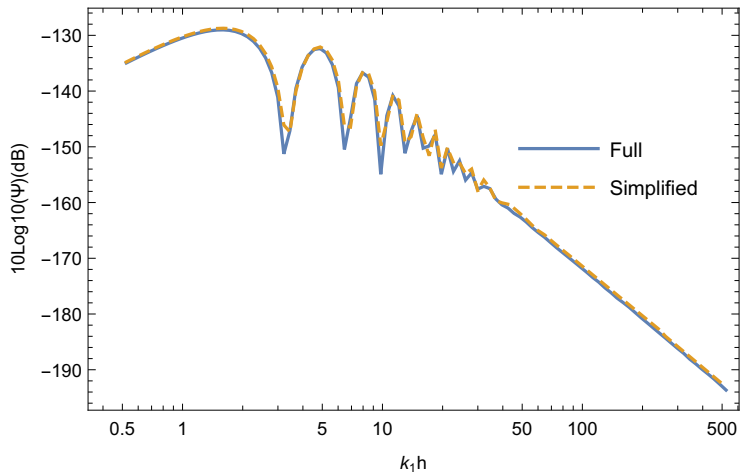}
    \caption{Comparison of predicted TE noise spectra from the full and
    simplified models: $M = 0.1$, $h = 5$, $\delta = 1.01$, $r = 30$, $\theta =
20^\circ$ and $y = 0$.}
    \label{fig:TEComparison20}
\end{figure}

Figure~\ref{fig:TEComparison20} shows the two predicted spectra when $\theta =
20^\circ$. The agreement continues to be very good over the entire frequency
range of interest. Although the two spectra are virtually identical, the
computational costs in obtaining them are, as those observed in the LE noise
problem, strikingly different: while the full solution demands an hour for
computing a single spectrum, the simplified model on average only needs a few
milliseconds. In summary, the approximated solution serves as an efficient
model for the TE noise problem. More importantly, the simplified TE noise model
can be easily implemented and the computation is very robust, while the
numerical integration in the full solution is prone to error due to the
non-smooth behaviour of the integrand.

\section{Comparison with experiments}
Results from these simplified models can be directly compared with experimental
data. Due to the similar nature of approximation, it suffices to focus on the
leading-edge model. We choose to compare with the recent experimental results
reported in \citet{Ayton2019} and use the sawtooth serration as an example. 

The experiment was carried out in the acoustic wind tunnel at the Institute of
Sound and Vibration of Southampton University. The test facility features a
$8\text{ m} \times 8 \text{ m} \times 8\text{ m}$ anechoic chamber and a
low-speed wind tunnel with a nozzle of $150 \text{ mm} \times 450\text{ mm}$.
More details on the test facility can be found in \citet{Ayton2019}. Flat
plates with serrated leading edges were placed $150\text{ mm}$ downstream from
the nozzle exit. The serrations had a wavelength of $25\text{ mm}$, but the
half root-to-tip amplitude was varied from $6.25\text{ mm}$ to $25\text{ mm}$.
Therefore, the corresponding $h$ was varied from $1/4$ and $1$.

\begin{figure}[!htpb]
    \centering
    \includegraphics[width=0.8\linewidth]{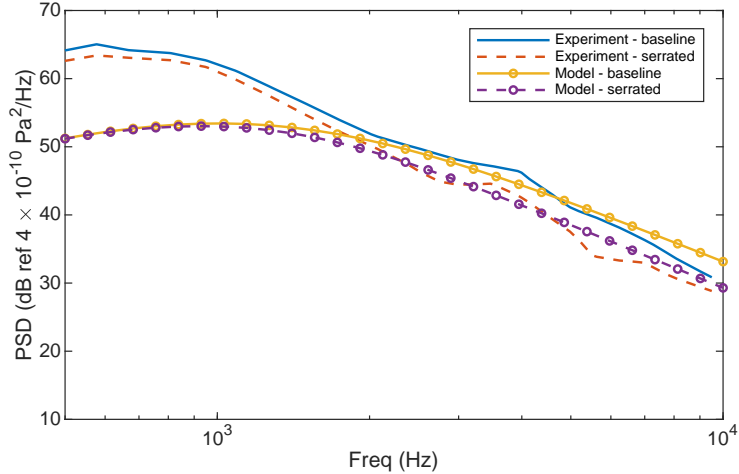}
    \caption{Comparison of the noise spectra between model prediction and
    experimental measurements when $h = 1/4$.}
    \label{fig:c=1}
\end{figure}

\begin{figure}[!htpb]
    \centering
    \includegraphics[width=0.8\linewidth]{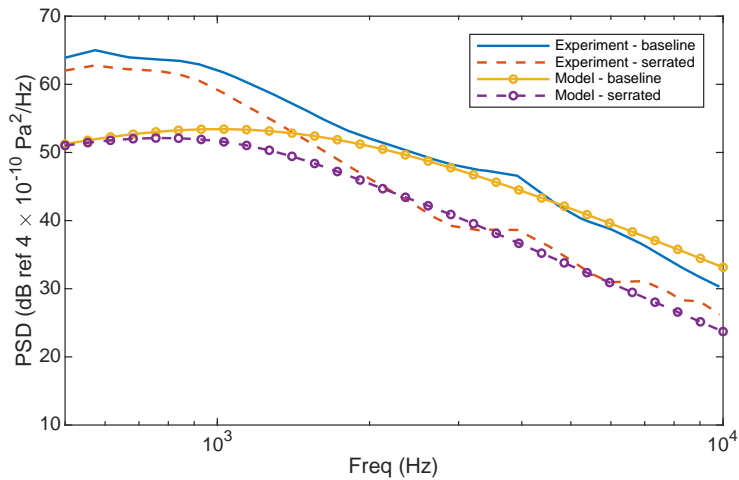}
    \caption{Comparison of the noise spectra between model prediction and
    experimental measurements when $h = 1/2$.}
    \label{fig:c=2}
\end{figure}

\begin{figure}[!htpb]
    \centering
    \includegraphics[width=0.8\linewidth]{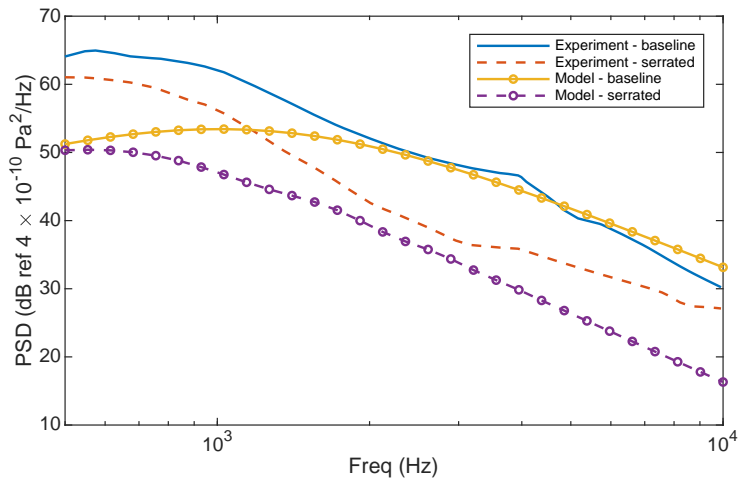}
    \caption{Comparison of the noise spectra between model prediction and
    experimental measurements when $h = 1$.}
    \label{fig:c=4}
\end{figure}

Free stream turbulence was generated by a rectangular grid of $630\text{ mm}
\times 690 \text{ mm}$ inside the contraction section located $75\text{ cm}$
upstream from the nozzle exit. The dimensionless turbulence spectrum
$\Pi_l(\omega, k_2)$ was characterised using the Liepmann model, i.e.
\begin{equation}
    \Pi_l(\omega, k_2) =  \frac{3 \mathrm{TI}^2 L_t^2}{4\pi} 
    \frac{L_t^2 (k_1^2 + k_2^2)}{(1 + L_t^2(k_1^2+k_2^2))^{5/2}},
    \label{equ:Liepmann}
\end{equation}
where $\mathrm{TI}$ and $L_t$ were, as defined in
section~\ref{subsec:LEComparison}, the turbulence intensity and streamwise
integral length scale, respectively. With equation~\ref{equ:Liepmann},
equation~\ref{equ:psdFarfieldSimplified2} can be quickly evaluated and the
results can be compared with the noise spectra obtained in the experiment.
These are presented in figures~\ref{fig:c=1} to \ref{fig:c=4}. To have an
intuitive understanding of the frequency and amplitude, noise spectra are shown
in their dimensional forms. 

We start comparing the model and experimental results for the short serration,
i.e. for $h = 1/4$. The far-field noise spectra are presented in
figure~\ref{fig:c=1}. Both the serrated and baseline ($h = 0$) spectra are
shown. It is well known that in leading-edge noise experiments low-frequency
noise is dominated by jet noise. Therefore, we do not make a direct comparison
for frequencies less than $2000$~Hz. Good agreement is achieved, however, in
the frequency range of $2000$ to $10000$~Hz for the baseline spectra. This
shows that the simplified model works well for straight edges. In addition, the
model predicts that a noise reduction of around $3$ dB can be achieved by using
the short serration of $h=1/4$. The experimental data agree with such a
prediction very well. 

Figure~\ref{fig:c=2} shows the comparison for $h = 1/2$. As can be seen, using
the longer serration results in a higher noise reduction of up to $8$ dB in the
experiment. The model can capture this change accurately and the resulting
spectrum for the serrated edge agrees very well with that observed in the
experiment. This is not surprising given that the simplified model agrees with
the full solution to a high degree of accuracy.

Figure~\ref{fig:c=4} shows the comparison for the long serration, i.e. for $h =
1$. We expect an even greater noise reduction due to the use of serrations, and
this is confirmed by the experiment, which shows a noise reduction of up to
more than $10$ dB. However, the predicted noise reduction is around $15$ dB.
The model therefore underpredicts the noise emitted by flat plates with long
leading-edge serrations. However, this is not due to the failure of the
simplified model, but rather due to neglecting contribution of trailing-edge
noise, which occurs in the experiment and begins to dominate for when
leading-edge noise is sufficiently suppressed. It has been shown by
\citet{Ayton2019} that adding in a trailing-edge noise model to the prediction
results in good agreement with the experimental spectrum, however, as we are
concerned with verifying the simplified models in this section, we have
excluded the contribution of TE noise here. Figures~\ref{fig:c=1} and
\ref{fig:c=4} show that the simplified model serves as an accurate
approximation to the full solution, which is both computationally expensive and
error-prone. The simplified model overcomes these two issues and is therefore
both efficient and robust.

\section{Conclusion and future work}
\label{sec:Conclusion}
This paper develops rapid noise prediction models for serrated leading and
trailing edges. This is based on the fact that high order modes are cut-off and
adjacent modes do not interfere in the far field except at sufficiently high
frequencies, so the infinite-interval integral involving two infinite sums may
be replaced by just one straightforward sum. The resulting models take
particularly concise forms when the serration is sufficiently narrow such that
$\bar{k} < \pi$ (or more roughly $\bar{k}<2\pi$) is satisfied in the frequency
range of interest. In practice this condition may afford further relaxation. A
comparison of these simplified models to the full analytical solutions shows
that the obtained models serve as excellent approximations over the entire
frequency range of interest. 

The leading-edge noise model is compared with experimental results for sawtooth
serrations of various root-to-tip amplitudes. Good agreement is achieved for
both $h = 1/4$ and $h = 1/2$. Deviation occurs for $h = 1$ but this is due to
the contribution of trailing-edge noise to the total noise observed in the
experiment, and the simplified model continues to approximate the full solution
with a great degree of accuracy. 

The models developed in this paper are robust, efficient, and can be easily
implemented. For example, a typical noise spectrum can be obtained within a few
milliseconds using these models, while the it takes hours to evaluate the
original full solutions. The efficiency and robustness would allow parametric
optimization studies to be performed quickly, which is important at the design
stage of many applications. 

%This paper only shows LE and TE noise spectra
%obtained at a fixed observer location and for a fixed Mach number. The
%dependence of these spectra on the observer location and Mach number will be
%studied as part of our future work.

\section*{Acknowledgement}
The authors wish to gratefully acknowledge the financial support
provided by the EPSRC under the grant number EP/P015980/1.

\bibliography{cleanRef}

\end{document}